%% file: 2025_IECON.tex
\documentclass[conference, hidelinks]{IEEEtran}
\IEEEoverridecommandlockouts
\usepackage{cite}
\usepackage{amsmath,amssymb,amsfonts}
\usepackage{algorithmic}
\usepackage{graphicx}
\usepackage{textcomp}
\usepackage{xcolor}

\usepackage{orcidlink}
\usepackage{flafter} % Placing floats after their definition
\usepackage{hyperref} % autoref
\usepackage{subcaption}
\usepackage{makecell}

\definecolor{UStuttDarkBlue}{RGB}{0,81,158}
\definecolor{UStuttDarkOrange}{RGB}{228,175,52}

\usepackage{tikz}
\usetikzlibrary{calc,fit, positioning,arrows.meta,spy}
\tikzset{>={Latex[width=2mm,length=2mm]}} % more visible default arrow heads
\tikzstyle{block} = [draw=black, fill=white, rectangle, align=center, minimum height=1em, minimum width=2em, font = {\footnotesize}]
\tikzstyle{smallblock} = [draw=black, fill=white, rectangle, align=center, minimum size=0.6cm, font = {\footnotesize}]
\tikzstyle{sum} = [draw, circle, node distance=1cm]
\tikzstyle{circ} = [draw=black, fill=white, circle, align=center, minimum size=0.4cm]
\tikzstyle{input} = [coordinate]
\tikzstyle{output} = [coordinate]
\tikzstyle{punkt} = [circle,fill,inner sep=0pt,minimum size=5pt]
\tikzstyle{gray dotted} = [draw=gray, fill=white, rectangle, align=center, line width=1pt, dash pattern=on 1pt off 4pt on 6pt off 4pt, inner sep=3mm, rounded corners, font = {\small}]
\tikzstyle{blueblock} = [draw=UStuttDarkBlue, fill=white, rectangle, align=center, minimum height=1em, minimum width=2em, line width =1.5pt, font = {\footnotesize}]
\tikzstyle{orangeblock} = [draw=UStuttDarkOrange, fill=white, rectangle, align=center, minimum height=1em, minimum width=2em, line width =2pt, font = {\footnotesize}]

\usepackage{siunitx} 
\DeclareSIUnit{\Bit}{Bit}

\def\BibTeX{{\rm B\kern-.05em{\sc i\kern-.025em b}\kern-.08em
    T\kern-.1667em\lower.7ex\hbox{E}\kern-.125emX}}
\begin{document}

\title{Lubricant Temperature Observer\\for Gearboxes in Industrial Robots
	\thanks{Funded by the Deutsche Forschungsgemeinschaft (DFG, German Research Foundation) – 443677015\\
		Data set available online:
		\url{https://doi.org/10.18419/DARUS-5015}
	}
}

\author{
	\IEEEauthorblockN{Christian J.\,E. Bauer\IEEEauthorrefmark{1} \orcidlink{0009-0002-7979-3877},
	Markus Neher\IEEEauthorrefmark{1},
	Valentin Kamm\IEEEauthorrefmark{1} \orcidlink{0000-0003-4748-5415},\\
	Lukas Steinle\IEEEauthorrefmark{1} \orcidlink{0000-0002-6035-0067},
	Armin Lechler\IEEEauthorrefmark{1} \orcidlink{0000-0002-4073-1487},
	Alexander Verl\IEEEauthorrefmark{1} \orcidlink{0000-0002-2548-6620}}
	\IEEEauthorblockA{\IEEEauthorrefmark{1}\textit{Institute for Control Engineering of Machine Tools and Manufacturing Units~(ISW)},\\
		University of Stuttgart, 70174 Stuttgart, Germany\\
		E-mail: christian.bauer@isw.uni-stuttgart.de}
}

\maketitle

\begin{abstract}
The position and path accuracy of industrial robots is strongly affected by temperature. In particular, temperature-dependent friction and altered hysteresis and transmission errors in the gearboxes are the main causes of these deviations. Since standard industrial robots lack temperature sensors, it is proposed to estimate the gearbox lubricant temperature using observer structures. It is planned to apply the proposed lubricant temperature observer to provide the input for temperature-dependent error compensation for gearboxes in industrial robots. First, a two-state heat transfer model describing the heat transfer from the lubricant through the gearbox housing to the environment is described and identified. It is integrated into a Luenberger observer and a Kalman filter with a correction of the lubricant temperature estimation based on measurements from a housing temperature sensor. Temperature measurements were performed in an experimental setup with a cycloidal gear filled with grease, which is common for industrial robots. Due to its changing viscosity and changing thermal conductivity at different shear rates, it shows a less uniform thermal behavior than gear-oil. For the validation measurements, the developed observers showed a mean absolute error of 1.3 K.
\end{abstract}

\begin{IEEEkeywords}
Data-driven modeling, Gears, Industrial robots.
\end{IEEEkeywords}

\section{Introduction}
\begin{figure}
	\centering
	\begin{subfigure}{0.655\columnwidth}
		\includegraphics[scale=1.0]{"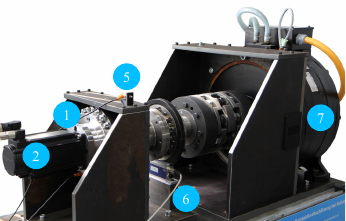"}
	\end{subfigure}
	\begin{subfigure}{0.325\columnwidth}
		\includegraphics[scale=1.0]{"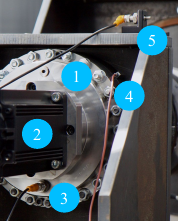"}
	\end{subfigure}
	\caption{Experimental setup~(presented in~\cite{mesmerModelingIdentificationHysteresis2022}): 1)~cycloidal drive, 2)~input-side motor, 3)~lubricant temperature sensor, 4)~housing temperature sensor, 5)~environment temperature sensor, 6)~torque sensor, 7)~load-side motor.}
	\label{fig:experiment-setup}
\end{figure}

The position accuracy of industrial robots is significantly affected by thermal changes.
For a standard industrial robot such as the ABB~IRB~2400, position errors of~\qty{0.441}{\mm} due to temperature changes can be measured at the tool center point~\cite{yinRealtimeThermalError2014}.
\textsc{Heisel} et~al. showed the influence of temperature on the position accuracy for SCARA, tripod, and six-axis kinematics.
For a single robot joint an angular error of~\qty{0.0006}{\degree\per\kelvin} was measured~\cite{heiselThermalBehaviourIndustrial1997}.
The drives are the main source of heat in industrial robots~\cite{sigronCompensationGeometricBacklash2024} due to the thermal power generated by friction.
In previous works, a compensation approach was introduced, accounting for temperature-dependent hysteresis and friction~\cite{mesmerInvestigationCompensationHysteresis2023}.
Since a temperature difference of~\qty{50}{\K} results in a change of~\qty{40}{\percent} to~\qty{60}{\percent} in the friction parameters~\cite{martonNovelApproachDeal2010}, a temperature-dependent friction model is required.
In~\cite{bauerModelingLoadVelocity2025}, the authors proposed a state-dependent transmission error compensation to further improve position accuracy.
Both approaches currently depend on sensors to measure the temperature of the gearbox lubricant.
Standard industrial robots are not equipped with gearbox temperature sensors~\cite{bittencourtExtendedFrictionModel2010, bittencourtModelingIdentificationWear2011, martonTemperatureDependentFriction2012, carlsonModelingIdentificationPosition2015a}.
Therefore, an observer approach to estimate the gearbox lubricant temperature was developed and validated in this work.
Temperature measurements were carried out with an experimental setup incorporating a single robot joint~(\autoref{fig:experiment-setup}).

\subsection{State of the Art}
\label{ssec:StateOfTheArt}
Current approaches to gear temperature estimation originate from research in temperature-dependent friction modeling.
Since friction is assumed to be temperature-dependent on the one hand, and friction is considered to be the main thermal input to the gearboxes on the other hand, the two fields are closely related.
\textsc{Bittencourt} et~al. extended a friction model for robot joints to include load- and temperature-dependent effects.
Robot gearboxes filled with oil were used in experiments to ensure a uniform temperature measurement by the lubricant temperature sensor.
An exponential description of the thermal effects on viscous friction was applied in~\hbox{\cite{bittencourtExtendedFrictionModel2010, bittencourtModelingIdentificationWear2011, bittencourtStaticFrictionRobot2012}}, which is related to the Vogel-Fulcher-Tammann equation~\cite{seetonViscosityTemperatureCorrelation2006}.
In~\cite{bittencourtModelingExperimentDesign2014}, the approach was extended to detect the wear of the modeled gear based on a deviation of the friction from the corresponding model.
\textsc{Márton} et~al. applied a friction model that describes the thermal influence via an internal, non-measurable state~\cite{martonNovelApproachDeal2010}.
This approach with a state corresponding to the liquefying of the lubricant under the influence of shear was introduced by \textsc{Batista} and \textsc{Carlson}~\cite{carlsonConstitutiveRelationFriction1996, batistaBifurcationsSteadySliding1998}.
The model was integrated with a viscous friction observer for robot gearboxes to detect changes in the lubricant consistency and wear and it was validated in simulation.
While the influence of temperature on viscous friction is assumed to show exponential behavior, the influence of thermal effects on the viscosity was modeled linearly~\cite{martonOnlineLubricantHealth2011}.
In~\cite{martonTemperatureDependentFriction2012}, the authors introduced a model based on the energy balance that describes the heat transfer from the lubricant over the gearbox housing to the environment.
It was integrated into a Kalman filter with the environment temperature and the product of velocity and friction torque as inputs to estimate the lubricant temperature.
This temperature estimation was used to detect faults based on changes in the friction parameters.
Tests were performed using a gear pair with a reduction ratio of~\num{3} and standard gear oil in the gearbox.
Temperatures in between approximately~\qty{22}{\degreeCelsius} and~\qty{32}{\degreeCelsius} were considered.
A similar two-state heat transfer model was used by~\textsc{Legnani} et~al.~\cite{legnaniModellingEvaluationFriction2019} and \textsc{Pagani} et~al.~\cite{paganiEvaluationModelingFriction2020} without an observer structure.
\textsc{Carlson} et~al. suggested a simpler model with a single state that considers only environment and gearbox temperature.
In their contribution, the approach was evaluated experimentally on two industrial robots.
The grease in the gearboxes was replaced by oil~\cite{carlsonModelingIdentificationPosition2015a}.
A similar model was used by \textsc{Simoni} et~al. to provide the input temperature to a third-order polynomial friction model.
The model was experimentally validated in an industrial robot with different gearboxes filled with oil and grease~\cite{simoniFrictionModelingTemperature2015}.
\textsc{Legnani} et~al. used the same robot to validate an alternative gearbox temperature model incorporating the duty cycle of the joint drive~\cite{legnaniNewFrictionModel2016}.
\textsc{Simoni} et~al. extended their approach to include the Stribeck effect~\cite{simoniInclusionTemperatureFriction2017} and evaluated it in a polishing task~\cite{simoniUseTemperatureBased2017}.
The single-state heat transfer model was embedded in an observer to estimate the initial gearbox temperature.
It showed a convergence time of a few minutes~\cite{simoniUseTemperatureBased2017}.
Increasing the observer gain led to an convergence in less than a second at cost of an oscillating and noisy estimation~\cite{simoniModellingTemperatureJoint2019}.
\textsc{Shimada} described a single-state model based on Newton's law of cooling for a lead-screw drive.
The model incorporates a disturbance observer to estimate the friction force as input~\cite{shimadaTemperatureEstimationLead2021}
\textsc{Nevmerzhitskiy} et~al. estimated the gearbox temperature based on measurements from the motor temperature sensor in a KUKA~KR10 and a correction term derived from a friction estimation during specific periodic trajectories~\cite{nevmerzhitskiyFrictionModelIndustrial2019}.
\textsc{Pagani} et~al. extended the two-state heat transfer model with an additional input for the thermal power from the motor and a third state for an optional temperature sensor placed nearby on the robot structure~\cite{paganiInfluenceHeatExchanges2020}.
A detailed analytical model of thermal lubrication characteristics for a cycloidal drive was proposed by \textsc{Han} et~al. for simulations~\cite{hanAnalysisThermalLubrication2024}.
\textsc{Paschold} et~al. derived an analytical thermal model of oil-lubricated gearboxes with multiple states based on the thermal network method~\cite{pascholdEfficiencyHeatBalance2020, pascholdCalculatingComponentTemperatures2022}.

\subsection{Contribution}
In this paper, a lubricant temperature observer for robot gearboxes based on the two-state heat transfer model proposed by \textsc{Márton} et~al. in~\cite{martonTemperatureDependentFriction2012} was developed.
In the state of the art, this approach was only elaborated for a small gearbox and comparably low temperatures, while the thermal behavior of industrial-scale gearboxes has not been addressed.
In this contribution, the thermal properties of a cycloidal drive with an \num{185}:\num{1} reduction ratio were investigated and modeled.
The experimental setup used for the temperature measurements represents a base joint in a standard industrial robot for heavy payloads.
A lubricant temperature range in between approximately~\qty{20}{\degreeCelsius} and~\qty{50}{\degreeCelsius} was investigated.\\
As common for the joints of many industrial robots, it is lubricated with grease instead of gear oil.
Since grease shows a reduced circulation within the gearbox, it has reduced heat dissipation properties compared to oil~\cite{siewerinInfluencesLubricantSupply2021}.
Grease is pasty at room temperature and with increasing temperature it decreases its viscosity.
Unlike oil, grease is a non-Newtonian fluid and shows a decreasing viscosity with increasing shear strain~\cite{seetonViscosityTemperatureCorrelation2006}.
While in~\cite{martonTemperatureDependentFriction2012} an oil-lubricated gearbox was considered, the suitability of a two-state heat-transfer model for grease-lubricated gearboxes was not investigated yet.
The specific properties of grease are taken into account in the present work.\\
In this work, the accuracy of the model without an observer structure was compared to a Luenberger observer (not covered in~\cite{martonTemperatureDependentFriction2012}) and a Kalman filter using the measured housing temperature for correction.
In addition to the analysis in~\cite{martonTemperatureDependentFriction2012}, the reliability of the temperature observer was tested by evaluating the transient behavior after an initialization with an incorrect initial lubricant temperature estimate.
The lubricant temperature observer is the basis for a potential sensorless application of the friction model and the joint hysteresis compensation in~\cite{mesmerInvestigationCompensationHysteresis2023} as well as the state-dependent transmission error model in~\cite{bauerModelingLoadVelocity2025}.
Standard industrial robots are not equipped with gearbox lubricant temperature sensors.
However, housing temperature sensors, which are easy and cost-effective to integrate, can be used as a basis to estimate the lubricant temperature.
If the robot has been idle for several hours and all mechanical components have cooled down to the known environmental temperature, the approach works even without housing temperature sensors~\cite{hanAnalysisThermalLubrication2024}.

\section{Methods}
\begin{figure}
	% System model
	\centering
	\input{"images/observer_structure.tex"}
	\caption{Structure of the temperature observer approach.}
	\label{fig:BlockDiagramOverview}
\end{figure}
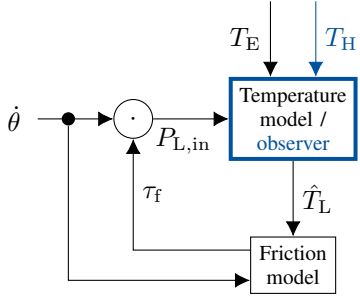

Temperature measurements with the single robot joint in the experimental setup in \autoref{fig:experiment-setup} were used to analyze its thermal behavior.
The two-state heat transfer model, the Luenberger observer and the Kalman filter were simulated and validated together with a temperature-dependent Stribeck curve as shown in~\autoref{fig:BlockDiagramOverview}.
The motor velocity~\(\dot{\theta}\) is measured and used together with the measured environment temperature~\(T_\mathrm{E}\) and housing temperature~\(T_\mathrm{H}\) to estimate the lubricant temperature~\(\hat{T}_\mathrm{L}\).
This estimated temperature and the motor velocity are the inputs to the friction model, which provides the friction torque~\(\tau_\mathrm{f}\).

\subsection{Experimental Setup}
\label{ssec:experimental_setup}
A Nabtesco~\hbox{\textit{RH380-N}} cycloidal drive with a reduction ratio of~\num{185} and~\qty{3724}{\N\m} rated output torque was used for the measurements in this paper.
Its transmission behavior was analyzed in~\cite{bauerModelingLoadVelocity2025}.
It is integrated into the experimental setup in \autoref{fig:experiment-setup}, which was introduced in~\cite{mesmerModelingIdentificationHysteresis2022}.
The gearbox is driven by a Bosch Rexroth~\textit{MSK070D} motor with~\qty{17.5}{\N\m} nominal torque.
Three PT100 temperature sensors are incorporated:
The lubricant temperature sensor is integrated into the gearbox opening for lubricant refill and a magnetic sensor is placed on the gearbox housing.
The environment temperature sensor is mounted on top of the gearbox mount with a thermally insulating angle bracket.
The signals from the sensors are digitized with a resolution of \qty{0.01}{\K} and a measurement uncertainty of $<\pm\qty{0.1}{\K}$ using Beckhoff~\hbox{\textit{EL3201-0010}}~\cite{manEtherCATTerminal1channel}.
Furthermore, a torque sensor and a load motor are integrated into the setup.
In this contribution, the load-free case was considered and therefore a~\qty{0}{\N\m} torque was commanded to the load motor. The combined inertia of load-side shaft, torque sensor, and coupling is~\qty{1.34}{\kg\m\squared} and the load-side motor has an inertia of~\qty{2.4}{\kg\m\squared}~\cite{manTechnicalDocumentationSynchronous}.\\
The gearbox is lubricated with Nabtesco~\hbox{\textit{RV GREASE LB00}}, a grease with a density of~\qty{0.87}{\g\per\cubic\cm}~\cite{manRVGREASELB002023}.
In~\cite{guanFrictionBehaviorWear2024}, the decrease in kinematic viscosity with increasing temperature and shear rate was measured for this non-Newtonian grease.
In the relevant temperature range of~\qty{25}{\degreeCelsius} to~\qty{50}{\degreeCelsius} it showed a kinematic viscosity of~\qty{200}{\milli\Pa\s} to~\qty{9000}{\milli\Pa\s} for a shear rate of~\(\num{10}\cdot\qty{1}{\per\s}\) to~\(\num{1000}\cdot\qty{1}{\per\s}\).

\subsection{Heat Transfer Model}
The temperature of the lubricant was derived from a two-state heat transfer model as described in~\cite{martonTemperatureDependentFriction2012}.
The thermal power injected by friction is considered as heat source for the lubricant.
It is transferred to the surrounding air through the steel housing of the gearbox.
The temperatures of the lubricant, the housing, and the environment are each assumed to be internally homogeneous.
Thermal conduction is expected to be the dominant heat transfer effect in the system.
Heat transfer between motor shaft and gearbox and between output shaft and gearbox were neglected.
The specific properties of grease as a non-Newtonian fluid are also neglected by the model assumption.\\
The thermal power injected into the lubricant~\(P_\mathrm{L,in}\) is defined as the product
\begin{align}
	P_\mathrm{L,in} = \bigl|\tau_\mathrm{f} \cdot \dot{\theta}\bigr| \label{eq:P_Lin}
\end{align}
of friction torque~\(\tau_\mathrm{f}\) and motor velocity~\(\dot{\theta}\).
The lubricant dissipates the thermal power
\begin{align}
	P_\mathrm{L,out} = K_\mathrm{LH} (T_\mathrm{L} - T_\mathrm{H})
\end{align}
to the housing, which is defined by the heat transfer coefficient from lubricant to housing~\(K_\mathrm{LH}\) and the difference between lubricant temperature~\(T_\mathrm{L}\) and housing temperature~\(T_\mathrm{H}\). The thermal power input to the housing
\begin{align}
	P_\mathrm{H,in} = P_\mathrm{L,out}
\end{align}
is equal to the lubricants thermal power output, since losses were not taken into account.
Analogous to~(\ref{eq:P_Lin}), the thermal power output of the housing
\begin{align}
	P_\mathrm{H,out} = K_\mathrm{HE}(T_\mathrm{H} - T_\mathrm{E})
\end{align}
 was derived from the heat transfer coefficient from housing to environment~\(K_\mathrm{LE}\) and the environment temperature~\(T_\mathrm{E}\).
 Inserting the derived power equations into the heat balance equations for the accumulated thermal power of the lubricant
 \begin{align}
 	P_\mathrm{L,acc} = P_\mathrm{L,in} - P_\mathrm{L,out} = C_\mathrm{L} \cdot	\dot{T}_\mathrm{L}
 \end{align}
 with the lubricant heat capacity~\(C_\mathrm{L}\) and the accumulated thermal power of the housing
 \begin{align}
 	P_\mathrm{H,acc} = P_\mathrm{H,in} - P_\mathrm{H,out} = C_\mathrm{H} \cdot	\dot{T}_\mathrm{H}
 \end{align}
with the housing heat capacity~\(C_\mathrm{H}\) leads to the model equations:
\begin{align}
	\dot{T}_\mathrm{L} = \frac{1}{C_\mathrm{L}} \left( \tau_\mathrm{f} \cdot \dot{\theta} - K_\mathrm{LH} (T_\mathrm{L} - T_\mathrm{H}) \right)~\text{and} \label{eq:dot_T_L}
\end{align}
\begin{align}
	\dot{T}_\mathrm{H} = \frac{1}{C_\mathrm{H}} \left (K_\mathrm{LH}(T_\mathrm{L} - T_\mathrm{H}) - K_\mathrm{HE}(T_\mathrm{H} - T_\mathrm{E}) \right)\text{.} \label{eq:dot_T_H}
\end{align}
These first-order differential equations can be transformed into a state-space model of the form
\begin{align}
	\dot{\boldsymbol{x}} &=
	\begin{bmatrix}
		-\frac{K_\mathrm{LH}}{C_\mathrm{L}} & \frac{K_\mathrm{LH}}{C_\mathrm{L}} \\
		\frac{K_\mathrm{LH}}{C_\mathrm{H}} & -\frac{K_\mathrm{LH} + K_\mathrm{HE}}{C_\mathrm{H}}
	\end{bmatrix}
	\boldsymbol{x}
	+
	\begin{bmatrix}
		\frac{1}{C_\mathrm{L}} & 0 \\
		0 & \frac{K_\mathrm{HE}}{C_\mathrm{H}}
	\end{bmatrix}
	\boldsymbol{u}\text{,} \nonumber \\
	\boldsymbol{y} &=
	\begin{bmatrix}
		0 & 1
	\end{bmatrix}
	\boldsymbol{x}\text{.}
\end{align}
It has the two states~\hbox{\(
\boldsymbol{x} = 
\begin{bmatrix}
 	T_\mathrm{L} \quad
	T_\mathrm{H}
\end{bmatrix}^\mathrm{T}
\)}, the inputs~\hbox{\(
\boldsymbol{u} = 
\begin{bmatrix}
	\tau_\mathrm{f} \cdot \dot{\theta} \quad
	T_\mathrm{E}
\end{bmatrix}^\mathrm{T}
\)}, and the output~\hbox{\(
\boldsymbol{y} = T_\mathrm{H}
\)}~\cite{martonTemperatureDependentFriction2012}.\\
The parameters of the heat transfer model were identified by applying the linear least squares method to individual measurements and taking the mean of the parameters over the identification measurements.
To avoid the need for the noisy derivatives~\(\dot{T}_\mathrm{L}\) and~\(\dot{T}_\mathrm{H}\) in the identification, the integrals
of~(\ref{eq:dot_T_L}) and~(\ref{eq:dot_T_H}) were used.
The identification was based on five identification measurements~\cite{manBauerReplicationDataLubricant} with the constant motor velocities~\(\dot{\theta}=i\cdot \qty{90}{\degree\per\s},~i\in\{0.6, 0.7, 0.8, 0.9, 1.0\}\).
The experimental setup was inactive for more than~\qty{12}{\hour} prior to all measurements to ensure that all components reached a temperature equilibrium.
The initial temperature of the lubricant was in between~\qty{19.3}{\degreeCelsius} and~\qty{21.5}{\degreeCelsius} and the measurements were stopped at~\qty{50}{\degreeCelsius}.\\
The identified state-space model was applied in a Luenberger observer and a Kalman filter.
While the model without observer structure only relies on the friction torque~\(\tau_\mathrm{f}\) and the environment temperature~\(T_\mathrm{E}\), the housing temperature~\(T_\mathrm{H}\) is included in the observers as measurement~(\autoref{fig:BlockDiagramOverview}).
The Luenberger observer matrix~\(\boldsymbol{L}\) and the covariance matrix of the process noise~\(\boldsymbol{Q}\) for the Kalman filter were chosen empirically to achieve a compromise between a short transient phase and low influence of measurement noise. The measurement noise variance~\(R\) was set as the measured variance of the lubricant temperature measurement in the idle state of the experimental setup.

\subsection{Friction Model}
The friction in the cycloidal gear was modeled with a temperature-dependent Stribeck curve
\begin{align}
	\tau_\mathrm{f}(\dot{\theta}, T) &= \left( \tau_{\mathrm{C}}(T) + (\tau_{\mathrm{s}}(T) - \tau_{\mathrm{C}}(T)) \cdot \mathrm{exp}\left(-\frac{\left|\dot{\theta}\right|}{v_{\mathrm{s}}(T)}\right)^{\delta_\mathrm{S}} \right) \nonumber \\
	&\cdot \mathrm{tanh} \left( \frac{1.5 \cdot \dot{\theta}}{v_\mathrm{b}}\right) + \sigma_\mathrm{v}(T) \cdot \dot{\theta}\text{,}
\end{align}
which depends on the Coulomb friction torque~\(\tau_{\mathrm{C}}\), the static torque~\(\tau_{\mathrm{s}}\), the Stribeck velocity~\(v_{\mathrm{S}}\), and the viscous friction factor~\(\sigma_\mathrm{v}\).
As suggested in~\cite{mesmerInvestigationCompensationHysteresis2023}, these parameters were described as quadratic equations
\begin{align}
	\tau_\mathrm{C} &= \tau_\mathrm{C,0} + \tau_\mathrm{C,1} \cdot T + \tau_\mathrm{C,2} \cdot T^2\text{,}\\
	\tau_\mathrm{S} &= \tau_\mathrm{S,0} + \tau_\mathrm{S,1} \cdot T + \tau_\mathrm{S,2} \cdot T^2\text{,}\\
	\sigma_\mathrm{v} &= \sigma_\mathrm{v,0} + \sigma_\mathrm{v,1} \cdot T + \sigma_\mathrm{v,2} \cdot T^2\mathrm{, and}\\
	v_\mathrm{S} &= v_\mathrm{S,0} + v_\mathrm{S,1} \cdot T + v_\mathrm{S,2} \cdot T^2\text{,}
\end{align}
depending on the temperature.
Using the tangens hyperbolicus instead of the signum function avoids a discontinuity at zero velocity, which can be problematic in an error compensation.
The parameter~\(v_\mathrm{b}\) determines the width of the transition between the sticking and slipping effect and the factor~\(\delta_\mathrm{S}\) is a shape parameter.
The friction parameters were identified with the Levenberg-Marquardt algorithm based on the friction measurements described in~\cite{mesmerInvestigationCompensationHysteresis2023}.

\section{Results}
The results of the identification process, the validation measurements, and the validation results are described in this section.

\subsection{Identification}
\label{ssec:Identification}
\begin{figure}
	\centering
	\includegraphics[scale=1]{"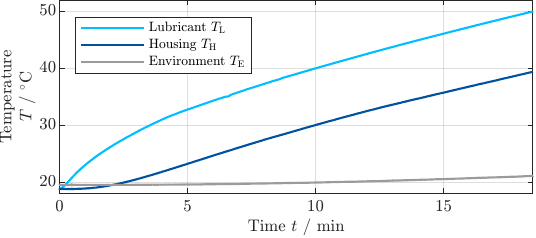"}
	\caption{Identification measurement for the velocity~\(\dot{\theta}=\qty{90}{\degree\per\s}\).}
	\label{fig:ident_1}
\end{figure}

The identification measurement for~\(\dot{\theta}=\qty{90}{\degree\per\s}\) is shown as an example in \autoref{fig:ident_1}.
It can be seen that the lubricant heated up exponentially in the first approximately~\qty{4}{\min}.
After that, it  heated up linearly. The housing temperature had a short transient phase before it followed the lubricant temperature with a near constant distance of~\qty{10}{\K}.
The environment heated up slowly during the measurement time.\\
The ranges of the identified model parameters for the individual measurements and the resulting model parameters are listed in~\autoref{tab:model-param}.\begin{table}[b]
	\caption{Identified Heat Transfer Model Parameters}
	\begin{center}
		\begin{tabular}{|c|c|c|c|}
			\hline
			\textbf{Parameter} & \textbf{Symbol}& \textbf{Identified range} & \textbf{Mean value}\\
			\hline
			\makecell{Lubricant heat\\capacity} & \(C_\mathrm{L}\) & \makecell{\(\qty{154.6376}{\joule\per\kelvin} \dots\) \\ \(\qty{179.0404}{\joule\per\kelvin}\)} & \qty{168.6251}{\joule\per\kelvin}\\
			\makecell{Housing heat\\capacity} & \(C_\mathrm{H}\) & \makecell{\(\qty{81.8903}{\joule\per\kelvin} \dots\) \\ \(\qty{109.7320}{\joule\per\kelvin}\)} & \qty{95.8339}{\joule\per\kelvin}\\
			\makecell{Heat transfer\\coefficient from\\lubricant to housing} & \(K_\mathrm{LH}\) & \makecell{\(\qty{0.1735}{\watt\per\kelvin} \dots\) \\ \(\qty{0.2518}{\watt\per\kelvin}\)} & \qty{0.2132}{\watt\per\kelvin}\\
			\makecell{Heat transfer coeffi-\\cient from housing\\to environment} & \(K_\mathrm{HE}\) & \makecell{\(\qty{0.0406}{\watt\per\kelvin} \dots\) \\ \(\qty{0.0813}{\watt\per\kelvin}\)} & \qty{0.0661}{\watt\per\kelvin}\\
			\hline
		\end{tabular}
		\label{tab:model-param}
	\end{center}
\end{table}
The Luenberger observer matrix was set to~\hbox{\(\boldsymbol{L}=
\begin{bmatrix}
	\num{0.1675}\cdot\qty{1}{\per\s} \quad
	\num{0.0368}\cdot\qty{1}{\per\s}
\end{bmatrix}^\mathrm{T}\)}
and the Kalman filter covariance matrix of the process noise to~\hbox{\(\boldsymbol{Q}=
 \mathrm{diag}\left(
 \begin{bmatrix}
 	\qty{0.4}{\K\squared} \quad
 	\qty{0.01}{\K\squared}
 \end{bmatrix}^\mathrm{T}\right)\)}.
A variance of the measurement noise of~\(R=\qty{7.2855e-5}{\K\squared}\) was measured.

\subsection{Validation}
\begin{figure}[t]
	\centering
	\begin{subfigure}{\columnwidth}
		\includegraphics[scale=1]{"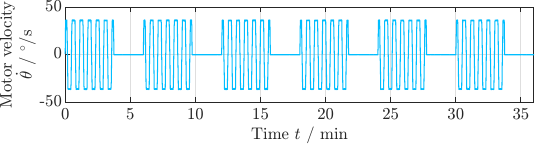"}
		\caption{Heat-up trajectory with~\(\dot{\theta}_\mathrm{max}=\qty{36}{\degree\per\s}\).}
		\vspace{0.25cm}
	\end{subfigure}
	\begin{subfigure}{\columnwidth}
		\includegraphics[scale=1]{"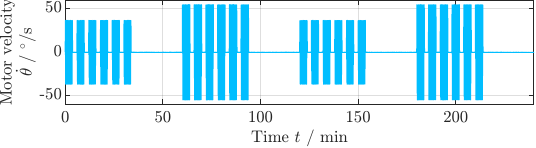"}
		\caption{Long-term measurement trajectory~\(\dot{\theta}_\mathrm{max}=\{\qty{36}{\degree\per\s}, \qty{54}{\degree\per\s}\}\).}
	\end{subfigure}
	\caption{Motor velocity trajectories for validation.}
	\label{fig:omega}
\end{figure}

\begin{figure*}[t]
	\centering
	\begin{subfigure}{0.495\textwidth}
		\begin{tikzpicture}[spy using outlines={circle, magnification=2, size=1.6cm, connect spies}]
			\node [anchor=south west, inner sep=0] (image) at (0,0) {\includegraphics[scale=1]{"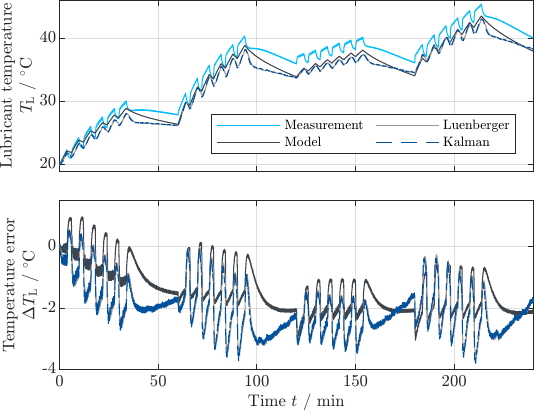"}};
			\spy on (4,6.04) in node [right] at (1.1,6.1);
			\node at (2.95,6.2) {\footnotesize 2×};
		\end{tikzpicture}
		\caption{Lubricant temperature.}
	\end{subfigure}
	\hfill
	\begin{subfigure}{0.495\textwidth}
		\includegraphics[scale=1]{"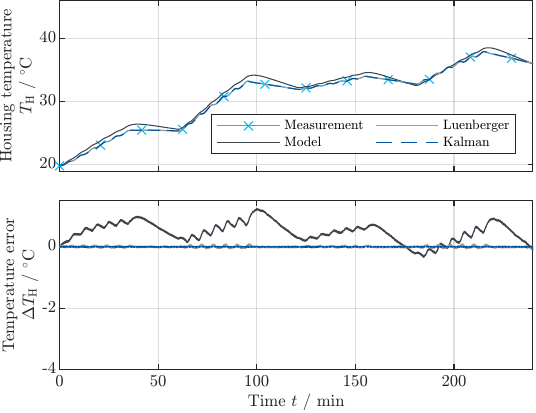"}
		\caption{Housing temperature.}
	\end{subfigure}
	\caption{Measurement and estimated values with long-term measurement trajectory~\(\dot{\theta}_\mathrm{max}=\{\qty{36}{\degree\per\s}, \qty{54}{\degree\per\s}\}\).}
	\label{fig:long_term_measurement_1}
\end{figure*}

\begin{figure*}
	\centering
	\begin{subfigure}{0.495\textwidth}
		\includegraphics[scale=1]{"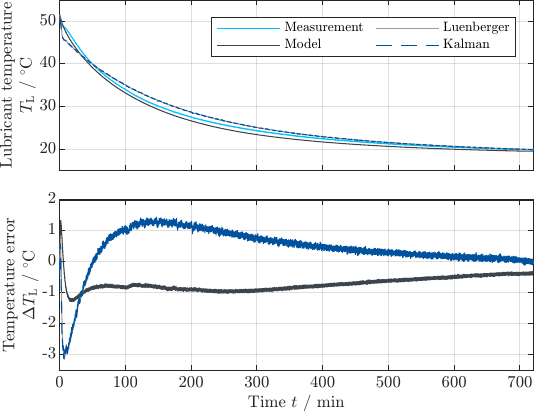"}
		\caption{Lubricant temperature.}
	\end{subfigure}
	\hfill
	\begin{subfigure}{0.495\textwidth}
		\includegraphics[scale=1]{"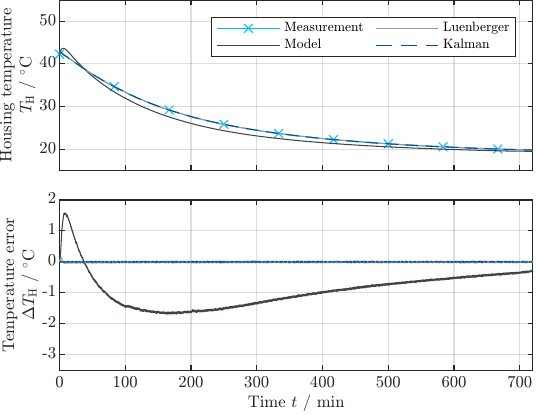"}
		\caption{Housing temperature.}
	\end{subfigure}
	\caption{Measurement and estimated values for cool-down.}
	\label{fig:cool_down_1}
\end{figure*}

\begin{figure}[t]
	\centering
	\includegraphics[scale=1]{"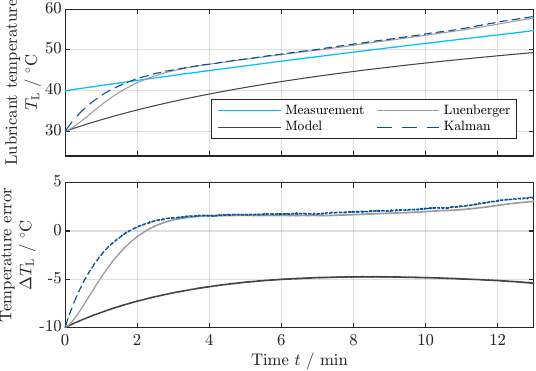"}
	\caption{Reliability test – Transient behavior for an incorrect initialization of the initial lubricant temperature estimate.}
	\label{fig:observer_initial_offset_ident_1}
\end{figure}

The model without observer structure, the Luenberger observer and the Kalman filter were simulated in MATLAB~/ Simulink with real measurement data as inputs using three validation scenarios:
Five heat-up trajectories, two long-term measurement trajectories combining several heat-up trajectories, and one cool-down measurement~\cite{manBauerReplicationDataLubricant}.
The heat-up trajectories alternated six times between the motor positions~\hbox{\(\theta=\pm\qty{720}{\degree}\)} with different maximum velocities~\hbox{\(\dot{\theta}_\mathrm{max}=i\cdot \qty{90}{\degree\per\s},~i\in\{0.2, 0.4, 0.6, 0.8, 1.0\}\)} as visualized in~\autoref{fig:omega}~(a).
The measurements were started at a lubricant temperature of approximately~\qty{20}{\degreeCelsius} and took~\qty{36}{\min}.
In the two long-term measurement trajectories two heat-up trajectories with~\hbox{\(\dot{\theta}_\mathrm{max}=\{\qty{36}{\degree\per\s}, \qty{54}{\degree\per\s}\}\) and~\(\dot{\theta}_\mathrm{max}=\{\qty{72}{\degree\per\s}, \qty{18}{\degree\per\s}\}\)} were combined and repeated twice~(\autoref{fig:omega}~(b)).
Also started with a lubricant temperature around~\qty{20}{\degreeCelsius}, they ran for about~\qty{4}{\hour}.
To analyze the cool-down behavior, the temperatures were measured with the input motor disabled for~\qty{12}{\hour} starting from a lubricant temperature of approximately~\qty{50}{\degreeCelsius}.\\
The maximum errors and the mean absolute errors~(MAE) for all validation measurements are listed in~\autoref{tab:errors}.
In \autoref{fig:long_term_measurement_1}, the model and observer performance are analyzed for the long-term measurement trajectory~\hbox{\(\dot{\theta}_\mathrm{max}=\{\qty{36}{\degree\per\s}, \qty{54}{\degree\per\s}\}\)}.
In general, the model without correction via housing temperature measurement has a smaller error between estimated and measured lubricant temperature compared to the observers.
All lubricant temperature estimates were slightly lower than the measurement.
The observers fitted the housing temperature, while the model overestimated the housing temperature.
The model fitted the lubricant temperature gradients better during the heat-up phases, but the observers fitted the gradient better during the cool-down phase when the gearbox was not moving.
The Luenberger observer had a slightly lower error than the Kalman filter in the lubricant temperature estimation, while the Kalman filter estimated the housing temperature more accurately.\\
To validate the behavior during cooling down for~\qty{12}{\hour}~(\autoref{fig:cool_down_1}), the observers were initialized with correct starting values.
Both observers estimated a steeper lubricant temperature decrease than it was measured for the first~\qty{50}{\min}.
Afterwards, they overestimated the measured value until they converged to the correct lubricant temperature after~\qty{12}{\hour}.
The model initially assumed a less steep gradient, but underestimated the temperature throughout the measurement.\\
To test the reliability of the temperature observers, the transient behavior of the model and observer estimates after an incorrect initialization of the initial lubricant temperature estimate was evaluated: The estimation for the identification measurement with the constant motor velocity~\(\dot{\theta}=\qty{90}{\degree\per\s}\) was started after reaching~\qty{40}{\degreeCelsius}, while the initial lubricant temperature was set to~\qty{30}{\degreeCelsius}~(\autoref{fig:observer_initial_offset_ident_1}).
While the model had an error of~\qty{-5}{\K} after~\qty{13}{\min}, the Luenberger observer had an error of~\qty{3.1}{\K} and the Kalman filter an error of~\qty{3.4}{\K}.
Both observers showed a transient phase for the first~\qty{3.2}{\min}, where the Kalman filter converged faster towards the measured value than the Luenberger observer.\begin{table}
	\caption{Errors Over Validation Measurements}
	\begin{center}
		\begin{tabular}{|c|c|c|}
			\hline
			& \textbf{Maximum error} & \textbf{MAE}\\
			\hline
			Model & \qty{3.4389}{\K} & \qty{0.8427}{\K}\\
			Luenberger observer & \qty{3.6098}{\K} & \qty{1.3073}{\K}\\
			Kalman filter & \qty{3.8232}{\K} & \qty{1.2956}{\K}\\ 
			\hline
		\end{tabular}
		\label{tab:errors}
	\end{center}
\end{table}

\section{Discussion}
The chosen two-state heat transfer model structure has been proven to be a valid approach for estimating the lubricant temperature for grease-lubricated robot gearboxes.
Moreover, its suitability for the application in a Luenberger observer and a Kalman filter has been shown.
While the model is able to fit the lubricant temperature during the heat-up phases, the model cannot reproduce the behavior during the idle phases.
The rapid drop in the lubricant measurement seen in~\autoref{fig:long_term_measurement_1} cannot be explained by cooling of the lubricant alone.
It is assumed that the lubricant is evenly distributed during the motion phases and accumulates at the bottom of the gearbox in the idle phase.
Additionally, the lubricant viscosity increases with decreasing temperature~\cite{seetonViscosityTemperatureCorrelation2006}, which could lead to an uneven distribution of the lubricant in the gearbox.
For non-Newtonian fluids such as grease, the thermal conductivity increases with shear rate~\cite{shinThermalConductivitySuspensions2000, sohnMicroconvectiveThermalConductivity1981, sohnHeatTransferEnhancement1984, charunyakornForcedConvectionHeat1991, leeShearRateDependent1997}.
This effect might explain the wide range of identified parameters for the different identification velocities~(in \autoref{ssec:Identification}) and the changing behavior in the idle state.
The results show, that the common use of grease instead of oil in robot gears increases the complexity of the lubricant temperature modeling problem.
The lubricant temperature estimates of the observers get corrected towards the housing temperature measurement.
Therefore, they show a better fit of the gradient of the unmodeled effects during the idle phases while their absolute value is less accurate.
Since the intended application is to increase position and path accuracy, as long as the estimation has a small error when starting a new trajectory, a correct estimation of the measured temperature behavior during the idle phases is not necessary.
The proposed observer structure has no feedback other than the calculated friction.
Therefore, differences between the provided simulation results based on  measurements and an execution of the observer on the experimental setup are not expected.
The stability of the proposed observer structure is ensured by the underlying model structures:
An overestimation of the lubricant temperature leads to an underestimation of the friction (due to the decreasing viscous friction with increasing temperature~\cite{mesmerInvestigationCompensationHysteresis2023}).
The underestimated friction causes a reduced temperature gradient estimation in the next iteration.
The opposite effect is present for an underestimation.
While the model without observer structure provides a more accurate temperature estimation for correct initial lubricant temperature values, the observers are able to correct an incorrect initial temperature estimate.
This test demonstrates the reliability of the temperature observers and their ability to recover from an initialization with wrong initial lubricant temperature estimates.
The Kalman filter converges faster to the measured temperature, but the Luenberger observer shows a smaller maximum error.
A small reduction in the values of the process noise covariance matrix of the Kalman filter~\(\boldsymbol{Q}\) could result in a Kalman filter that relies less on the housing temperature measurement, thus reducing the error.
In the intended use of the temperature observer to provide the input for the hysteresis compensation in~\cite{mesmerInvestigationCompensationHysteresis2023} and the transmission error compensation in~\cite{bauerModelingLoadVelocity2025}, a temperature accuracy of less than~\qty{5}{\K} is assumed to be appropriate.
Overall, the maximum error of the Kalman filter of~\qty{3.8}{\K} and the MAE of~\qty{1.3}{\K} is therefore small enough to provide a significant improvement in position accuracy.
If the initial lubricant temperature estimate is correct, the model without observer structure and without temperature sensor shows a lower maximum error of~\qty{3.4}{\K} and a MAE of~\qty{0.8}{\K}.

\section{Outlook}
In further steps to improve the proposed approach, a wider range of velocities could be used during parameter identification.
In this paper, only the load-free case was evaluated.
A possible extension to consider the additional power input due to load-dependent friction (e.~g. as in~\cite{mesmerInvestigationCompensationHysteresis2023}) has to be investigated.
Furthermore, the heat transfer model could be adapted to include the heat transfer from the motor, which was neglected in the presented approach.
Since industrial servomotors are usually equipped with a temperature sensor, its usability as an additional or alternative observer measurement should be considered.
The effectiveness of the motor temperature sensor as the only measurement for the observer structures is unclear because it is mounted in the coils at the distal end of the motor and is separated from the motor shaft by the air gap between stator and rotor.
Alternatively, the motor temperature sensor can be used to estimate the initial temperature after a long idle phase when all components reached equilibrium temperature.
For industrial robots that are not used at night, a possible scenario could be the initial lubricant temperature estimation based on the motor temperature sensor combined with the model without observer structure.
Finally, the combined use with the hysteresis compensation in~\cite{mesmerInvestigationCompensationHysteresis2023} and the transmission error compensation in~\cite{bauerModelingLoadVelocity2025} has to be validated.
Afterwards, an identification and evaluation of the temperature observer in an industrial robot is planned.

\addtolength{\textheight}{-14.5cm}	% This command serves to balance the column lengths

\section*{Data Availability}
The temperature measurements presented in this paper are available on DaRUS, the data repository of the University of Stuttgart: \url{https://doi.org/10.18419/DARUS-5015}

\bibliographystyle{IEEEtran}
\bibliography{IEEEabrv, 2025_IECON, 2025_IECON_manual}

\end{document}

%% file: images/observer_structure.tex
\begin{tikzpicture}
	\node[circ] (x) {$\cdot$};
	\node[punkt, left=0.5cm of x] (input) {};
	\node[blueblock, right=1cm of x] (temp) {Temperature\\model / \\ \textcolor{UStuttDarkBlue}{observer}};
	\node[block, below=1cm of temp] (friction) {Friction\\model};
	
	\draw[->] ($(x.west)+(-1,0)$) -- node[left, xshift=-0.6cm] {\(\dot{\theta}\)} (x);
	\draw[->] (x) -- node[below, xshift=-0.1cm] {\(P_\mathrm{L,in}\)} (temp);
	\draw[->] ($(friction.west)+(0,0.2)$) -| node[pos=0.75, right] {\(\tau_\mathrm{f}\)} (x);
	\draw[->] (temp.south) -- node[right] {\(\hat{T}_\mathrm{L}\)} (friction.north);
	\draw[->] ($(temp.north)+(-0.3,1)$) --  node[left] {\(T_\mathrm{E}\)} ($(temp.north)+(-0.3,0)$);
	\draw[->, color=UStuttDarkBlue] ($(temp.north)+(0.3,1)$) --  node[right] {\(T_\mathrm{H}\)}($(temp.north)+(0.3,0)$);
%	\draw[->] ($(temp.east)+(0,0.3)$) -- node[right, xshift=0.6cm] {\(\hat{T}_\mathrm{H}\)} ($(temp.east)+(1,0.3)$);
%	\draw[->] ($(temp.east)+(0,-0.3)$) -- node[right, xshift=0.6cm] {\(\hat{T}_\mathrm{L}\)} ($(temp.east)+(1,-0.3)$);

	\draw[->] (input) |- ($(friction.west)+(0,-0.2)$);
\end{tikzpicture}